\documentclass{article}
\usepackage{emulateapj,pstricks}
\usepackage{graphicx}
\newcommand{\ASCA}{{\it ASCA} }
\newcommand{\ROSAT}{{\it ROSAT} }
\newcommand{\EUVE}{{\it EUVE} }
\newcommand{\BeppoSAX}{{\it BeppoSAX} }
\newcommand{\RXTE}{{\it RXTE} }
\newcommand{\citep}{\cite}
\newcommand{\citet}{\cite}
\nonstopmode

\righthead{FUKAZAWA ET AL.}
\lefthead{DETECTION OF EXCESS HARD X-RAY EMISSION FROM HCG62}
\submitted{Preparing for the Astrophysical Journal.}

\begin{document}

\title{Detection of Excess Hard X-ray Emission from the Group of Galaxies HCG62}
\renewcommand{\baselinestretch}{1.0}

\author{
Yasushi Fukazawa,$^{1,2}$ Kazuhiro Nakazawa,$^2$ Naoki Isobe,$^2$ Kazuo Makishima,$^2$\\
Kyoko Matsushita,$^{2,3}$ Takaya Ohashi,$^4$ and Tsuneyoshi Kamae,$^{1,2}$ 
}

\affil
{
$^1$ Department of Physical Sciences, Graduate School of Science,
              Hiroshima University, 1-3-1 Kagamiyama, Higashi-Hiroshima-shi, Hiroshima 739-8256\\
\it   E-mail(YF) fukazawa@hirax6.hepl.hiroshima-u.ac.jp\\
$^2$ Department of Physics, Graduate School of Science, University of Tokyo, 7-3-1 Hongo, Bunkyo-ku, Tokyo 113-0033\\
$^3$ Max-Planck-Institut f\"ur Extraterrestrische Physik, Giessenbachstra\ss e, D-85748 Garching, Germany\\
$^4$ Department of Physics, Faculty of Science, Tokyo Metropolitan University, Minami-Ohsawa, Hachioji, Tokyo, 192-0397
}

\begin{abstract}

From the group of galaxies HCG62,
we detected an excess hard X-ray emission in energies above $\sim 4$ keV with \ASCA.
The excess emission is spatially extended up to $\sim10'$ from the group center,
and somewhat enhanced toward north.
Its spectrum can be represented by
either a power-law of photon index 0.8--2.7,
or a Bremsstrahlung of temperature $>6.3$ keV.
In the 2--10 keV range, the observed hard X-ray flux, 
$(1.0\pm0.3)\times10^{-12}$ erg cm$^{-2}$ s$^{-1}$,
implies a luminosity of $(8.0\pm2.0)\times10^{41}$ erg s$^{-1}$
for a Hubble constant of 50 km s$^{-1}$ Mpc$^{-1}$.
The emission is thus too luminous to be attributed to X-ray binaries in the member galaxies.
We discuss possible origin of the hard X-ray emission.
\end{abstract}


\section{Introduction}

Clusters of galaxies are thought to have released 
a large amount of dynamical energy in their initial collapse phase.
During their subsequent evolution, starburst-driven winds, 
cluster mergers, radio galaxies, and random galaxy motions
may have supplied additional heating energy to the intracluster space.
Presumably, these processes have generated energetic particles
(e.g. \citet{kang,takizawa}),
as evidenced by diffuse synchrotron radio emission from some clusters.

Such energetic particles are expected to produce nonthermal X-rays as well,
by Compton-boosting cosmic microwave background (CMB) photons.
Long searches for such effects among galaxy clusters have recently revealed two candidates;
the excess soft X-ray emission detected with \EUVE \citep{lieu,mittaz,bowyer},
and the spectral hard X-ray tail observed with \BeppoSAX \citep{fusco,kaastra}.
However, the exact nature of these emission components remain unclear.

Groups of galaxies are the poorest class of galaxy clusters.
Their thermal emission is limited to energies below $\sim 5$ keV,
because the temperature of their hot intragroup medium is about 1 keV
(e.g. \citet{mulchaey,fuka96}).
Therefore, they allow us to search for nonthermal X-ray emission,
even with instruments operating below an energy of $\sim 10$ keV.
Here we report the detection of excess hard X-ray emission 
from the group of galaxies HCG62 with the \ASCA GIS \citep{ohashi,makishima}.
We employ 90\% confidence limits throughout this paper, 
and use the Hubble constant of 50 km s$^{-1}$ Mpc$^{-1}$.
Solar abundances refer to \citet{anders}.

\section{Observations and Data Reduction}

With a redshift of 0.0137 \citep{hickson},
HCG62 is one of the nearest Hickson compact galaxy groups.
It was observed twice with \ASCA;
on 1994 January 14--15 in a single pointing,
and on 1998 January 13--17 in 4 pointings to cover the whole group region.
The GIS was operated in PH mode,
and the SIS in 2CCD FAINT mode in 1994.
We do not use the SIS data taken in 1998,
because of the insufficient field of view of 1 CCD mode employed at that time.
After an appropriate gain correction,
we co-added all the available data from different sensors, chips, and pointings,
separately for the GIS and the SIS.
The livetime is $\sim 30$ ksec for the 1994 observation, 
and $\sim 20$ ksec for each of the 4 pointings of the 1998 observation.
The total GIS livetime thus amounts to 110 ksec.

For our purpose, 
it is important to accurately subtract the GIS background,
which consists of cosmic X-ray background (CXB)
and intrinsic detector background (IDB).
We first summed data of the \ASCA Large Sky Survey \citep{ueda},
conducted in 1993 December and 1994 June over blank sky fields, 
with a total exposure time of 233 ksec.
Then, after \citet{ikebe}, we excluded regions in the GIS images
where count rate exceeds those from surrounding regions by $\geq$2.5$\sigma$.
This eliminates faint sources with the
2--10 keV flux $>8\times10^{-14}$ erg s$^{-1}$ cm$^{-2}$.

We next corrected the IDB level of each pointing individually
for its gradual increase by 2--3\% per year,
and for its random day-by-day fluctuation by 6--8\% \citep{ishisaki}.
For this purpose, we derived three GIS spectra, denoted $S(E)$, $B(E)$, and $N(E)$,
from the on-source data, the blank sky data prepared as above, 
and night earth data, respectively.
They were accumulated over an annulus of radius $13'-25'$ from the GIS field center
and in the 6--10 keV energy range,
to ensure that $S(E)$ is free from the HCG62 emission,
and that the CXB is relatively minor compared to the IDB in $S(E)$ and $B(E)$.
Then, assuming that the IDB spectrum and its radial profile are both constant,  
we fitted $S(E)$ with a linear combination $B(E)+fN(E)$;
here $f$ is a free parameter,
and $fN(E)$ represents the secular IDB change
between the two epochs when $S(E)$ and $B(E)$ were acquired.
We have obtained $f=0.00$ and $f=0.08-0.12$
for the 1994 data and those of 1998, respectively, 
in agreement with the IDB long-term increase \citep{ishisaki}.
By analyzing various \ASCA data,
we also confirmed that this method can reproduce, to within 5\%,
the GIS background spectra and its radial profiles
acquired at any epoch over 1993--1999.

The SIS has a lower efficiency in the hard X-ray band, 
a shorter exposure time,
and a smaller field of view, than the GIS.
We therefore utilize the SIS spectrum only to determine 
the soft thermal emission from the intragroup medium.
We subtract the SIS background in a conventional way,
utilizing the archival SIS background set.

\section{Results}

To avoid the diffuse thermal emission with a typical 
plasma temperature of $kT\sim 1.0$ keV \citep{ponman,fuka98,davis}, 
we produced the GIS image of HCG62 in the hard 4.5--8 keV band,
as shown in figure 1.
There, we overlaid the 1.0--2.4 keV image as a measure of the thermal emission,
of which the brightness peak coincides in position with the group center to within $1'$.
The image reveals a hard X-ray emission,
which apparently extends up to $\sim10'$ from the group center.

Figure 2 shows the radial GIS count-rate profile in the energy band of 4.5--8 keV, 
centered on the soft X-ray brightness peak.
Also shown are the instrumental point-spread function (PSF),
and the profile of the estimated background.
Thus, the background level is well reproduced at larger radii within 5\%,
and the hard X-ray emission is more extended than the PSF,
detectable up to $10'$ from the group center.
As shown in the inset to figure 2,
the hard X-ray surface brightness is higher in the north region than in the south region.
Such a feature cannot be explained as a spill-over from the 1 keV thermal emission.
The hard X-ray brightness is not correlated with the galaxy distribution, either,
including emission line galaxies \citep{carvalho}.

The observed hard X-ray emission, though apparently extended,
could simply be a result of several hard point sources,
such as active galaxies, either related or unrelated to HCG62.
To answer this issue, we examined the archival \ROSAT image of HCG62,
and found four point sources with 0.1--2 keV fluxes of
$(4-8)\times10^{-14}$ erg s$^{-1}$ cm$^{-2}$ 
(shown in figure 1 as open squares)
at the locations where 
the hard-band GIS image actually exhibits possible enhancements 
with the implied 2--10 keV fluxes of $\sim 10^{-13}$ erg s$^{-1}$ cm$^{-2}$.
The flux ratio between \ASCA and \ROSAT indicates that 
the source spectra have a power-law shape of photon index $ \sim 1.5$.
We have accordingly excluded photons falling within $2.5'$ of these four sources.
In addition, in order to remove possible point-like sources at the central region of HCG62,
we excluded photons within $3'$ of the group center.
Then, the 4.5--8 keV GIS2+GIS3 flux from the on-source data
has become $3772 \pm 61$ photons over the radius of $3'-15'$,
compared to $3351 \pm 20$ expected from the background count rate.
The excess, $421 \pm 65$ counts,
well exceeds the $\sim$60 counts expected for the 1 keV thermal emission.
Thus, the presence of the extended excess hard X-ray emission is significant 
from the GIS imagery even after excluding possible point-source contamination.

In order to examine the excess hard X-ray emission through spectroscopy,
we have produced spectra over the radius of $3'-15'$,
by utilizing all the available data from the GIS 
and only the first pointing data from the SIS.
The regions within 2.5$'$ of the four point-like sources were again excluded.
We discarded the SIS data above 4 keV for the reason described before.
The obtained spectra are presented in figure 3.
We fitted them simultaneously,
by a single temperature plasma emission model \citep{rs} (R-S model) 
with solar abundance ratios \citep{fuka96,fuka98}, modified by photoelectric absorption.
As shown in table 1 and figure 3,
the model successfully reproduced the data in lower energies,
and the derived temperature and metallicity are consistent with those of \citet{fuka98}.
However, the model is not acceptable due to significant residuals 
seen in the GIS fit over energies of $> 3$ keV.
When we limit the energy band to $<2.4$ keV,
the fit becomes acceptable with a reduced chi square of 1.29
and the best-fit temperature of $0.95 \pm 0.05$ keV.
In contrast, when we use only the hard energy band above 2 keV,
the best-fit temperature increases to $2.1\pm0.3$ keV;
this is inconsistent with that indicated by the soft-band data,
and is much higher than the prediction from the galaxy velocity dispersion of
$\sim300$ km s$^{-1}$ \citep{mulchaey}.
These results reconfirm the presence of excess hard X-ray emission 
above the prediction of the thermal emission of temperature $\sim 1$ keV.
The results of \citet{finoguenov}, who reported a high temperature of
$>1.5$ keV around $5'$ of the group center, are also consistent with ours.

We refitted the whole-band spectra
by adding a Bremsstrahlung or a power-law component,
to represent the excess hard X-ray emission.
As summarized in table 1,
either modeling has given an acceptable joint fit to the \ASCA spectra.
The Bremsstrahlung temperature has been constrained as $>6.3$ keV,
while the power-law photon index $\alpha_{\rm ph}$ was found at $1.5_{-0.7}^{+1.2}$.
Although $\alpha_{\rm ph}$ can be as high as 2.7,
such a steep power-law forces the R-S component to have an extremely high metallicity.
When we fix the metallicity of the R-S component at 0.30 solar
which is typically found from clusters of galaxies,
the upper limit on $\alpha_{\rm ph}$ becomes 2.2.
Below, we utilize this limit instead of the original one.
The normalization of the hard component does not change by more than 20\%,
if we use plasma emission codes other than the R-S code. 

The upper limit on narrow Fe-K line features at $\sim 6.6$ keV
is uninteresting, several keV in equivalent width.
The absorption column density cannot be constrained in any case,
and consistent with the Galactic value of
$2.9\times10^{20}$cm$^{-2}$ \citep{stark}.
The 2--10 keV X-ray flux and luminosity of the hard X-ray component
are $(1.0\pm0.3)\times10^{-12}$ erg cm$^{-2}$ s$^{-1}$
and $(8.0\pm2.0)\times10^{41}$ erg s$^{-1}$, respectively,
regardless of the choice between the two modelings.
This amounts to about 20\% of the  0.5--10 keV thermal-component luminosity 
of $4.9\times10^{42}$ erg s$^{-1}$ (Fukazawa 1997).

Although we have carefully estimated the background,
it is still important to examine
to what extent our results are affected by possible background uncertainties.
To see this, we intentionally increased the IDB background level by 5\%,
and found that the 2--10 keV flux and luminosity of the hard component
becomes $(5.5-8.4)\times10^{-13}$ erg s$^{-1}$ cm$^{-2}$
and $(4.3-6.5)\times10^{41}$ erg s$^{-1}$, respectively.
Thus, the hard emission remains statistically significant.

\section{Discussion}

From the galaxy group HCG62, we have detected the excess X-ray emission
with a very hard spectrum, which extends up to more than 10$'$ from the
group center and somewhat enhanced at the north region.
Although its surface brightness is only $\sim$20\% of that of the GIS background,
we have confirmed its reality through careful analysis.
Below, we discuss the origin and nature of this phenomenon.

An immediate possibility is collection of binary X-ray sources 
in the member galaxies of HCG62.
However, based on the total optical luminosity of HCG62 
($\sim1\times10^{11}L_{\rm \odot}$; \citet{carvalho})
and the optical vs. X-ray luminosity correlation 
among elliptical galaxies \citep{matsushita},
this contribution is estimated to be at most $4\times10^{40}$ erg s$^{-1}$,
which is an order of magnitude short of the observed luminosity.
A second possibility is assembly of faint
active galactic nuclei (AGNs) in HCG62.
However, the optical evidence for AGNs in HCG62 is moderate \citep{carvalho},
and we have already subtracted such candidates based on the \ROSAT image.
Any remaining AGNs are estimated to contribute 
no more than 30--40\% of the total hard X-ray emission.
Yet another possibility is fluctuation of background faint sources.
Utilizing the log$N$--log$S$ relation in the 2--10 keV band \citep{ueda}, 
this contribution is estimated to be at most
$\sim2\times10^{-13}$ erg cm$^{-2}$ s$^{-1}$ over the radius of $3'-15'$,
which is again too low to explain the data.
From these considerations, we conclude
that the excess hard X-ray emission cannot be explained
by assembly of discrete hard X-ray sources,
whatever their nature be.

Considering the loose constraint on the Fe-K line,
the excess emission might be of thermal origin from very hot plasmas.
Actually, \citet{buote} described the \ASCA spectra of HCG62,
integrated over a radius of $0'$--$3'$,
by a two-temperature plasma model of $kT=0.7$ and $kT=1.5$ keV.
We have independently reconfirmed their results.
However, our spectra (Figure 3) that are accumulated
over the $3'$--$15'$ range requires a temperature of $> 6.3$ keV;
the original two-temperature model found by \citet{buote}
gives a very poor fit ($\chi^2/\nu = 1.66$).
Thus, thermal emission with an ``ordinary'' temperature cannot explain
the data. 
Insignificant detection of excess hard X-rays 
over a radius of $0'$--$3'$ might be due to poor photon statistics and 
complex spectra of thermal components \citep{fuka98} at the center region, 
spectral change of hard components, and so on.
There might still be plasmas much hotter than the escape temperature,
as is actually found in starburst galaxies (e.g. \citet{ptak}).
However, within $15'$ of HCG62,
there is no bright IRAS sources with the 60 $\mu$m flux exceeding 2 Jy
(SkyView at High Energy Astrophysics Science Archive Research Center).
We therefore conclude that the thermal interpretation of
the excess hard X-ray emission is unrealistic.

Given the difficulties with the discrete-source and thermal
interpretations of the diffuse hard X-ray emission,
we regard nonthermal interpretation as most promising.
One popular scenario of non-thermal X-ray production is inverse Compton scattering
of the CMB photons by relativistic electrons with Lorentz factor $\gamma\sim 10^{3-4}$,
as has been invoked to explain the excess hard X-ray emission 
from rich clusters (e.g. \citet{fusco}).
However, we cannot constrain the intragroup magnetic field in HCG62
due to lack of information on the diffuse radio flux.
If we assume a representative magnetic field intensity of 1 $\mu$G,
we would observe synchrotron radio emission with a flux density of $\sim$0.3 Jy. 
Since such a strong radio emission is not seen from HCG62 (SkyView; NVSS VLA image), 
the inverse-Compton interpretation holds for HCG62
only if its magnetic field is much weaker than 1 $\mu$G.
The reality of such a weak magnetic field is an open question, 
and in an apparent contradiction to the 
generally accepted inter-galactic field strengths of $\sim 1~\mu$G \citep{kronberg},
even though such a condition is suggested by the \BeppoSAX and \RXTE observations 
of some rich clusters of galaxies \citep{valinia,fusco}.

An alternative interpretation is nonthermal Bremsstrahlung between
the thermal gas and subrelativistic particles,
as proposed for the hard X-ray emission from Abell~2199,
of which diffuse radio flux is quite weak \citep{kempner}
like in the case of HCG62.
Let us assume for simplicity
that the nonthermal electrons have typical energies of 10--100 keV,
and their spatial density distribution is similar to that of the thermal intragroup gas.
Then, the nonthermal to thermal luminosity ratio
in the 0.5--10 keV band becomes $\sqrt{10}\alpha$,
where $\alpha$ is the density ratio of the nonthermal electrons to the thermal ones.
The observed luminosity ratio of $\sim 0.2$ implies $\alpha\sim$0.06,
indicating that the energy density of nonthermal electrons is
0.6--6 times as high as that of thermal electrons (depending on the spectrum).
If this is the case, the mechanism of such particle acceleration becomes an important issue.
In addition, the non-thermal pressure associated with such a particle population
would considerably increase the total mass of HCG62, and hence its dark-matter content,
estimated from the X-ray data.

The diffuse hard X-ray emission has been observed with \ASCA 
from some other galaxy groups as well \citep{fuka99}.
However, its prominence relative to the thermal X-ray emission appears
to scatter from object to object like rich clusters \citep{molendi},
with HCG62 one of the strongest case.
Although it is yet to be studied what makes such variety,
the hard X-ray emission might be related with transient phenomena such as mergers.

The authors are grateful to Dr. Takizawa for helpful discussions, and
to an anonymous referee for helpful comments.
This research was supported in part by the Grants-in-Aid for the 
Center-of-Excellence (COE) Research of the Ministry of Education, Science, 
and Culture in Japan (07CE2002).

\begin{table*}
\caption[]{Results of joint fitting of the GIS and SIS spectra of HCG62 with various models.}
\begin{center}
\begin{tabular}{cccccc}
\hline
\hline
$N_{\rm H}^{\dagger}$ & $kT^{\star}$  & abundance$^{\star}$  & norm1$^{\star}$ & $\alpha_{\rm ph}$ & normalization of bremsstrahlung \\
($10^{20}$cm$^{-2}$) & (keV) & (solar) & ($10^{17}$cm$^{-5}$)& $kT_{\rm bremss}$ & or powerlaw model\\
\hline
\multicolumn{6}{l}{Raymond-Smith ($\chi^2$/dof = 2.24)}\\
$<3.7$ & 1.02$\pm$0.03 & 0.15$\pm$0.02 & 7.2$\pm$0.6 & & \\
\hline
\multicolumn{6}{l}{Raymond-Smith + Bremsstrahlung ($\chi^2$/dof = 0.99)}\\
$<3.5$ & 0.95$\pm$0.05 & 0.18$\pm$0.05 & 5.9$\pm$1.0 & $kT_{\rm bremss}>6.3$ keV & $1.2_{-0.2}^{+0.7}\times10^{17}$ cm$^{-5}$ \\
\hline
\multicolumn{6}{l}{Raymond-Smith + Powerlaw ($\chi^2$/dof = 0.99)}\\
$<3.4$ & 0.96$_{-0.6}^{+0.3}$ & $>0.14$ & $5.6_{-3.5}^{+1.3}$ & $\alpha_{\rm ph}=1.5_{-0.7}^{+1.2}$ & $1.7_{-1.0}^{+5.7}\times10^{-4}$ c s$^{-1}$ cm$^{-2}$ keV$^{-1}$ \\
\hline
\multicolumn{6}{p{17cm}}{\small $\dagger$: Column density of photoelectric absorption in unit of $10^{20}$cm$^{-2}$.}\\
\multicolumn{6}{p{17cm}}{\small $\star$: Parameters of Raymond-Smith model.}\\
\end{tabular}
\end{center}
\end{table*}

\begin{figure}
\begin{center}
{\resizebox{120mm}{!}{\includegraphics*[width=12cm]{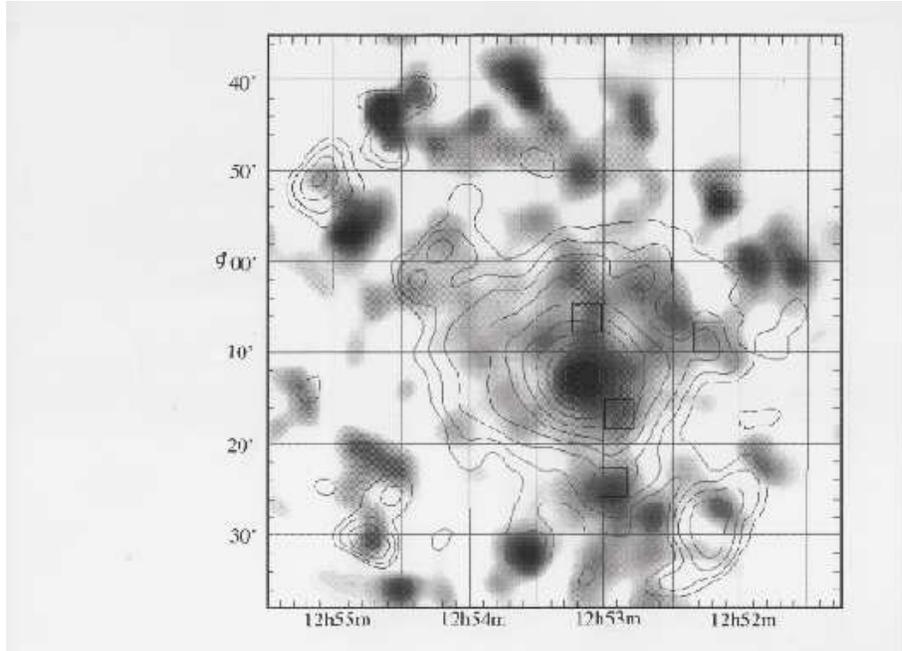}}}
\caption{
The background-subtracted GIS image of HCG62,
in 4.5--8 keV (gray scale) and 1.0--2.4 keV (contours).
Both images have been smoothed with a Gaussian filter of $\sigma=1'$,
and their scales are logarithmic.
Four squares are positions of the \ROSAT point sources
which would be also detected with the GIS.
}
\end{center}
\end{figure}

\begin{figure}
\begin{center}
\resizebox{83mm}{!}{\includegraphics[width=8cm]{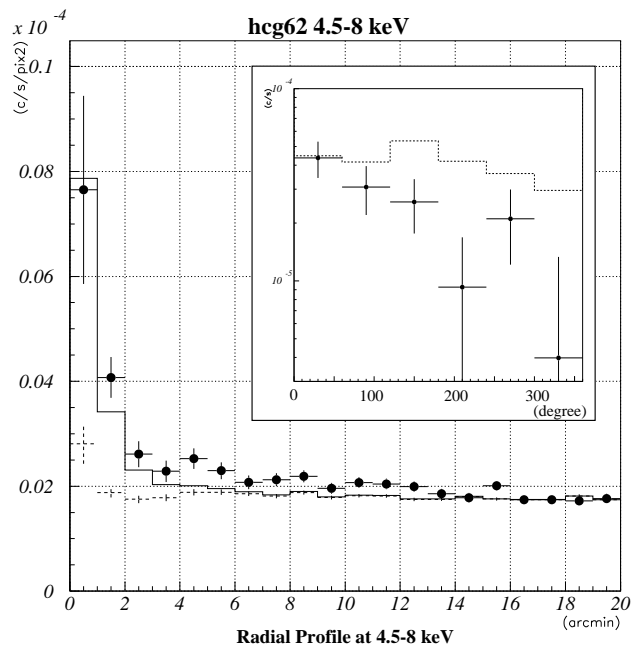}}
\caption{
The GIS radial count rate profile of HCG62 (crosses with filled circles) centered on the soft
 X-ray peak position, including the background. The dashed crosses
represent the estimated background, and the solid line shows 
the \ASCA XRT+GIS point spread function plus the estimated background.
The inset shows the background-subtracted azimuthal count rate profile in
 4.5-8 keV (crosses with filled circles) and 1.0--2.4 keV (dotted histogram).
The angle is defined counterclockwisely with the east being the origin.
}
\end{center}
\end{figure}

\begin{figure}
\begin{center}
\resizebox{120mm}{!}{\includegraphics[width=12cm]{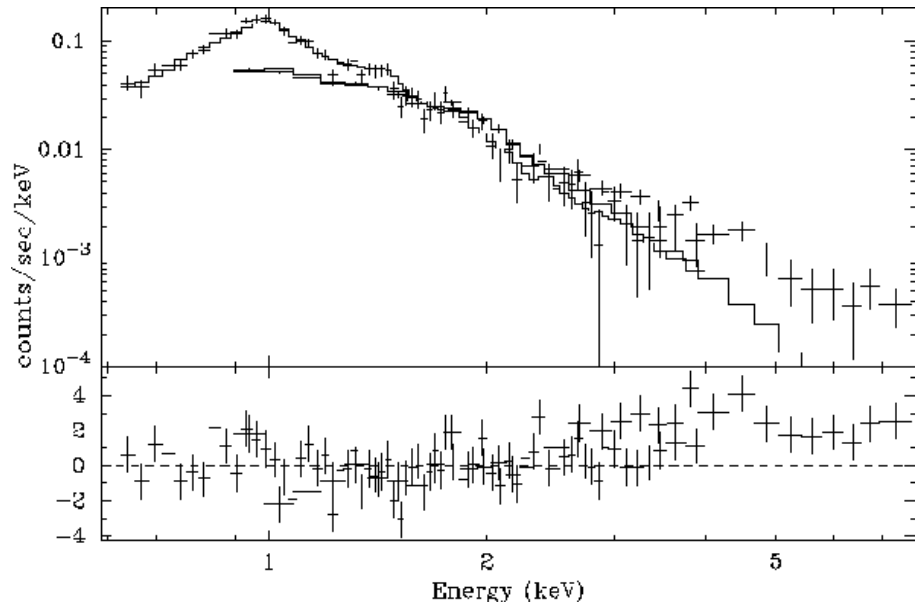}}
\caption{
The GIS+SIS simultaneous spectral fitting of HCG62 with the
 single temperature Raymond-Smith model. The cross points and solid lines represent
 the data and model, respectively.
The SIS data above 4 keV are discarded.
}
\end{center}
\end{figure}

\end{document}